\documentclass[pra,aps,showpacs,onecolumn]{revtex4}
\UseRawInputEncoding
\usepackage{amssymb}
\usepackage{amsmath}
\usepackage{mathtools}
\usepackage[dvipdfm]{hyperref}
\usepackage{hyperref}

\allowdisplaybreaks
\begin{document}

\title{$l_{1}$ norm of coherence is not equal to its convex roof quantifier}
\author{Jianwei Xu}
\email{xxujianwei@nwafu.edu.cn}
\affiliation{College of Science, Northwest A\&F University, Yangling, Shaanxi 712100,
China}


\begin{abstract}
Since a rigorous framework for quantifying quantum coherence was established
by Baumgratz et al. [T. Baumgratz, M. Cramer, and M. B. Plenio, Phys. Rev.
Lett. 113, 140401 (2014)], many coherence measures have been found. For a
given coherence measure $C$, extending the values of $C$ on pure states to
mixed states by the convex roof construction, we will get a valid coherence
measure $\overline{C}$, we call $\overline{C}$ the corresponding convex roof
quantifier of $C$. Whether $C=\overline{C}$ for a given coherence measure is
an important question. In this work, we show that for the widely used
coherence measure, $l_{1}$ norm of coherence $C_{l_{1}}$, it holds that $%
C_{l_{1}}\neq \overline{C_{l_{1}}}$.
\end{abstract}

\maketitle

\section{Introduction}
Coherence is a basic ingredient for quantum physics. Since a rigorous
framework for quantifying quantum coherence was established by Baumgratz et
al. \cite{BCP-2014-PRL} (we call this framework BCP framework), fruitful
results have been achieved both in theories and experiments (reviews see
\cite{Plenio-2016-RMP,Fan-2018-PhysicsReports}).

We first review the BCP framework as follows. For the $d$-dimensional
Hilbert space $H$, the coherence of the states in $H$ under BCP framework is defined with
respect to a fixed orthonormal basis $\{|j\rangle \}_{j=1}^{d}$ of $H.$ In
this work, when we say coherence, which is always with respect to this fixed
orthonormal basis $\{|j\rangle \}_{j=1}^{d}.$ A state $\rho $ is called
incoherent if and only if $\rho _{jk}=\langle j|\rho |k\rangle =0$ for any $%
j\neq k.$ A channel $\phi $ on $H$ is called incoherent if and only if $\phi
$ allows for a Kraus operator decomposition $\phi =\{K_{l}\}_{l}$ which
satisfying $\langle j|K_{l}\rho K_{l}^{\dagger }|k\rangle =0$ for any
incoherent state $\rho ,$ any $l$ and any $j\neq k$, and $%
\sum_{l}K_{l}^{\dagger }K_{l}=I_{d}$ with $I_{d}$ the identity operator on $%
H.$ We call such decomposition $\phi =\{K_{l}\}_{l}$ an incoherent
decomposition of $\phi $. A coherence measure $C$ for quantum states in $H$
should satisfy (C1)-(C4) below \cite{BCP-2014-PRL}.

(C1). Nonnegativity. $C(\rho )\geq 0$ for any $\rho ,C(\rho )=0$ if and only
if $\rho $ is incoherent.

(C2). Monotonicity. $C(\phi (\rho ))\leq C(\rho )$ for any $\rho $ if $\phi $
is incoherent.

(C3). Strong monotonicity. $\sum_{l}$tr$(K_{l}\rho K_{l}^{\dagger })C(\frac{%
K_{l}\rho K_{l}^{\dagger }}{\text{tr}(K_{l}\rho K_{l}^{\dagger })})\leq
C(\rho )$ for any $\rho $ if $\phi $ is incoherent and $\phi =\{K_{l}\}_{l}$
an incoherent decomposition of $\phi $.

(C4). Convexity. $C(\sum_{l}p_{l}\rho _{l})\leq \sum_{l}p_{l}C(\rho _{l})$
for any states $\{\rho _{l}\}_{l}$ and any probability distribution $%
\{p_{l}\}_{l}.$

A condition (C5) was proposed in Ref. \cite{Tong-2016-PRA}, and it was shown
that (C3)+(C4) imply (C2), (C2)+(C5) are equivalent to (C3)+(C4).

(C5). Additivity for direct sum states.
\begin{eqnarray}
C(p\rho _{1}\oplus (1-p)\rho _{2})=pC(\rho _{1})+(1-p)C(\rho _{2}), \label{eq1.1}
\end{eqnarray}
with $p\in \lbrack 0,1],\rho _{1},\rho _{2}$ any states.

Many coherence measures have been proposed, such as the relative entropy of
coherence $C_{\text{r}}$ \cite{BCP-2014-PRL}$,l_{1}$ norm of coherence $%
C_{l_{1}}$ \cite{BCP-2014-PRL,Yang-2016-PRL}$,$ coherence based on Tsallis
relative entropy $C_{T,\alpha }$ \cite{Rastegin-2016-PRA,Yu-2017-PRA,Yu-2018-SR,Xiong-2018-PRA}, robustness of
coherence $C_{\text{rob}}$ \cite{Adesso-PRL-2016,Adesso-PRA-2016}, geometric
coherence \cite{Adesso-2015-PRL} $C_{\text{g}}$, modified trace norm of
coherence $C_{\text{tr}}^{\prime }$ \cite{Tong-2016-PRA}, coherence weight
\cite{Bu-2018-PRA,Yao-2020-PRA}. Among them, $C_{\text{r}}$ and $C_{l_{1}}$
are widely used, they have the explicit expressions as
\begin{eqnarray}
C_{\text{r}}(\rho) &=&S(\rho _{\text{diag}})-S(\rho ),  \label{eq1.2}  \\
C_{l_{1}} (\rho) &=&\sum_{j\neq k}|\rho _{jk}|,   \label{eq1.3}
\end{eqnarray}
where $S(\rho )=-$tr$(\rho \log _{2}\rho )$ is the Von Neumann entropy of $%
\rho $, $\rho _{\text{diag}}$ is the diagonal part of state $\rho .$

There is a useful way of constructing some coherence measures by the convex
roof construction via the concave functions \cite%
{Yuan-2015-PRA,Guo-2015-PRA,Qi-2015-QIC}. This construction first specifies
the coherence for pure states, and then extends to mixed states by taking
the convex roof over all pure state decompositions of the given mixed state. Specifically, suppose $%
f(p_{1},p_{2},...,p_{d})$ is a nonnegative function on the $d$-dimensional
probability space, and also, $f(p_{1},p_{2},...,p_{d})$ is concave,
invariant under the index permutation of $\{j\}_{j=1}^{d}$, and $%
f(1,0,...,0)=0$. Then we can get the coherence measure $C_{f}$ induced by $f$
defined as
\begin{eqnarray}
C_{f}(|\psi \rangle \langle \psi |) &=&f(|\langle \psi |1\rangle
|^{2},|\langle \psi |2\rangle |^{2},...,|\langle \psi |d\rangle |^{2}),  \label{eq1.4}  \\
C_{f}(\rho ) &=&\min_{\{p_{j},|\psi _{j}\rangle \}}\sum_{j}p_{j}C_{f}(|\psi
_{j}\rangle \langle \psi _{j}|),   \label{eq1.5}
\end{eqnarray}%
where $|\psi \rangle $ is a normalized pure state, min runs over all pure
state decompositions of $\rho =\sum_{j}p_{j}|\psi _{j}\rangle \langle \psi
_{j}|$ with $\{p_{j}\}_{j}$ a probability distribution and $\{\psi
_{j}\}_{j} $ normalized pure states.

Evidently, there are many such functions $\{f\}$, and then there are many
corresponding coherence measures $\{C_{f}\}$ using the convex roof
construction. However, it is hard to calculate $C_{f}(\rho )$ in general
because of the definition of minimization.

There is another way to construct a new coherence measure via a given
coherence measure also using the convex roof construction. For a given
coherence measure $C$, define the coherence measure $\overline{C}$ as
\begin{eqnarray}
\overline{C}(\rho )=\min_{\{p_{j},|\psi _{j}\rangle \}}\sum_{j}p_{j}C(|\psi
_{j}\rangle \langle \psi _{j}|),   \label{eq1.6}
\end{eqnarray}
where min runs over all pure state decompositions of $\rho
=\sum_{j}p_{j}|\psi _{j}\rangle \langle \psi _{j}|.$ We see that $\overline{C%
}(\rho )=C(\rho )$ for any pure state $\rho $ and $\overline{C}(\rho
)=C(\rho )=0$ for any incoherent state $\rho $. We call $\overline{C}$ the
corresponding convex roof coherence measure (or coherence quantifier) of $C.$
It is shown in Ref. \cite{Yuan-2015-PRA} that $\overline{C_{\text{r}}}$ is a
valid coherence measure, i.e. $\overline{C_{\text{r}}}$ fulfills (C1)-(C4).
In fact, it can be proven in the similar way as in Ref. \cite{Yuan-2015-PRA}
that for any coherence measure $C$, $\overline{C}$ is indeed a coherence
measure, i.e. $\overline{C}$ fulfills (C1)-(C4). $\ \overline{C_{\text{r}}}$
was studied in Refs. \cite{Aberg-2006-arxiv,Yuan-2015-PRA,Yang-2016-PRL}, $\
\overline{C_{l_{1}}}$ was studied in Ref. \cite{Qi-2017-JPA}. Since $\overline{C%
}(\rho )=C(\rho )$ for any pure state, then
\begin{eqnarray}
\overline{\overline{C}}=\overline{C}.   \label{eq1.7}
\end{eqnarray}

An important question arises that whether $C=\overline{C}$. Note that here $%
C=\overline{C}$ means $C(\rho )=\overline{C}(\rho )$ for any state with any
dimension $d$, while we say $C\neq \overline{C}$ when there exists one state
$\rho $ such that $C(\rho )\neq \overline{C}(\rho )$. From the convexity of
coherence measure and the definition of $\overline{C},$ we see that
\begin{eqnarray}
C\leq \overline{C},   \label{eq1.8}
\end{eqnarray}
and $C=\overline{C}$ if and only if for any state $\rho $ there exists a
pure state decomposition $\rho =\sum_{j}p_{j}|\psi _{j}\rangle \langle \psi
_{j}|$ such that
\begin{eqnarray}
C(\rho )=\sum_{j}p_{j}C(|\psi _{j}\rangle \langle \psi _{j}|).   \label{eq1.9}
\end{eqnarray}
This means the pure state decomposition $\rho =\sum_{j}p_{j}|\psi
_{j}\rangle \langle \psi _{j}|$ realizes the state $\rho ,$ meanwhile
achieves the coherence $C(\rho ).$

We see that the coherence measure $C_{f}$ defined by Eqs. (\ref{eq1.4},\ref{eq1.5}) via the
function $f$ satisfies $C_{f}=\overline{C_{f}}.$ Since there is an infinite
number of such functions $\{f\},$ hence there exists an infinite number of
coherence measures satisfying $C=\overline{C}.$

The geometric coherence, $C_{g},$satisfies $C_{g}=\overline{C_{g}}.$ This
fact was pointed out in Ref. \cite{Plenio-2016-RMP}.
For $C_{\text{r}},$ it is shown that for qubit system \cite{Yuan-2015-PRA},
\begin{eqnarray}
\overline{C_{\text{r}}}(\rho )=h(\frac{1+\sqrt{1-r^{2}+z^{2}}}{2}),   \label{eq1.10}
\end{eqnarray}
where $h(x)=-x\log _{2}x-(1-x)\log _{2}(1-x)$ is the binary entropy, and we
have used the Bloch ball representation that any qubit state $\rho $ can be
expressed as $\rho =\frac{1}{2}(I_{2}+\overrightarrow{r}\cdot
\overrightarrow{\sigma })$ with $\overrightarrow{r}=(x,y,z)$ real vector
satisfying $r=\sqrt{x^{2}+y^{2}+z^{2}}\leq 1,\overrightarrow{\sigma }%
=(\sigma _{x},\sigma _{y},\sigma _{z})$ the Pauli matrices.
From Eq. ((\ref{eq1.2}), for qubit system, we have
\begin{eqnarray}
C_{\text{r}}(\rho )=h(\frac{1+z}{2})-h(\frac{1+r}{2}).   \label{eq1.11}
\end{eqnarray}
Figure 1(a) and Figure 1(b) show $\overline{C_{\text{r}}}(\rho )\geq C_{\text{r}}(\rho )$ and $\overline{C_{\text{r}}}(\rho )-C_{\text{r}}(\rho )$ for $%
0\leq z\leq r\leq 1.$ We see that $\overline{C_{\text{r}}}(\rho )-C_{\text{r}%
}(\rho )>0$ for most states. For example let $z=0,r=\frac{1}{\sqrt{2}},$ we
have $C_{\text{r}}(\rho )=1-h(\frac{1+\frac{1}{\sqrt{2}}}{2})<\overline{C_{%
\text{r}}}(\rho )=h(\frac{1+\frac{1}{\sqrt{2}}}{2}).$ This shows
\begin{eqnarray}
C_{\text{r}}\neq\overline{C_{\text{r}}}.   \label{eq1.12}
\end{eqnarray}
\begin{figure}[!ht]
\centering
  \begin{minipage}[!ht]{0.49\textwidth}
    \centering
    \includegraphics[width=1\textwidth,bb=10 0 270 170]{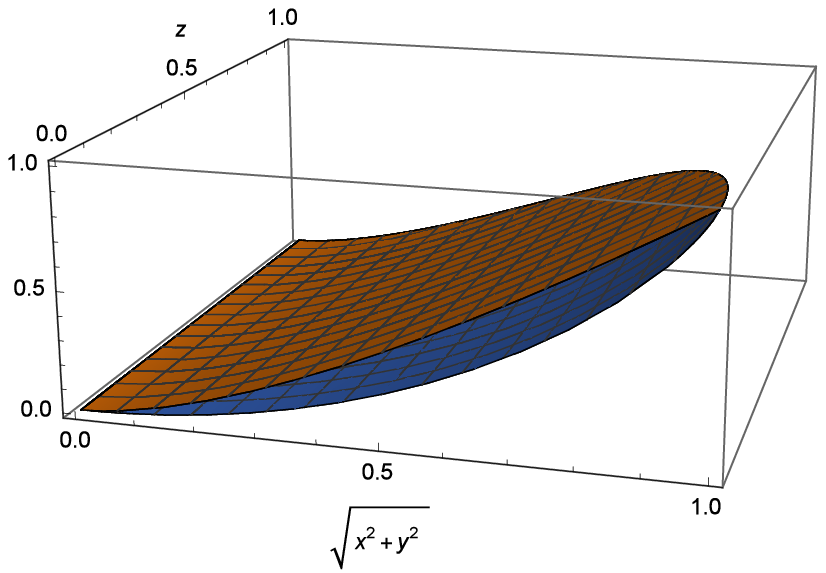}
    Fig.1(a):$\overline{C_{\text{r}}}\geq C_{\text{r}}$
  \end{minipage}
  \begin{minipage}[!ht]{0.4\textwidth}
    \centering
    \includegraphics[width=1\textwidth,bb=10 0 210 160]{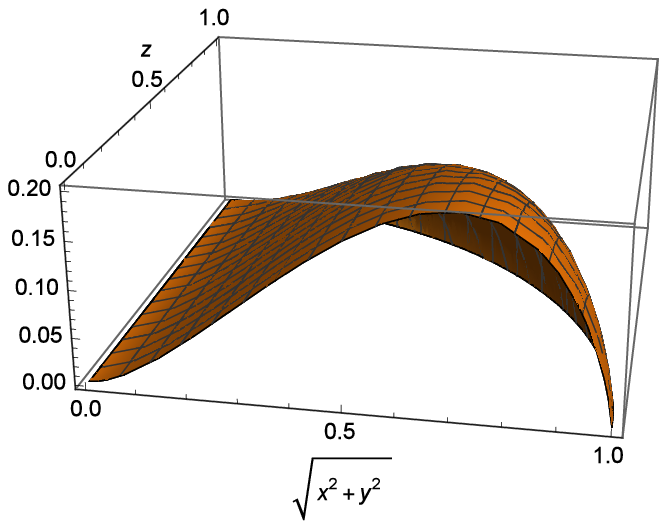}
    Fig.1(b):$\overline{C_{\text{r}}}-C_{\text{r}}$
  \end{minipage}
\end{figure}

In Ref. \cite{Qi-2017-JPA}, it is shown that
$\overline{C_{l_{1}}}=C_{l_{1}}\text{ for }d=2$,
whether $\overline{C_{l_{1}}}=C_{l_{1}}$ for $d>2$ is still an open question.
In this work, we show that $\overline{C_{l_{1}}}\neq C_{l_{1}}$. This work
is organized as follows. In section II we establish a criterion for $%
\overline{C_{l_{1}}}=C_{l_{1}}.$ In section III we show that $\overline{%
C_{l_{1}}}=C_{l_{1}}$ for $d=2.$ In section IV we show that $\overline{%
C_{l_{1}}}\neq C_{l_{1}}$ for $d=3.$ Section V is a brief summary.

\section{A criterion for $\overline{C_{l_{1}}}(\rho )=C_{l_{1}}(\rho )$}
We establish a criterion when a state $\rho $ satisfies $\overline{C_{l_{1}}}%
(\rho )=C_{l_{1}}(\rho ).$

\emph{Theorem 1.} For $d$-dimensional state $\rho =\sum_{j,k=1}^{d}\rho
_{jk}|j\rangle \langle k|$ with $\rho _{jk}=|\rho _{jk}|e^{i\theta _{jk}}$
its polar form, then $\overline{C_{l_{1}}}(\rho )=C_{l_{1}}(\rho )$ if and
only if there exists a pure state decomposition $\rho =\sum_{l}p_{l}|\psi
_{l}\rangle \langle \psi _{l}|$ such that
\begin{eqnarray}
\langle j|\psi _{l}\rangle \langle \psi _{l}|k\rangle =|\langle j|\psi
_{l}\rangle \langle \psi _{l}|k\rangle |e^{i\theta _{jk}},\forall j,k,l.   \label{eq2.1}
\end{eqnarray}

\emph{Proof.} When $j=k$, we have $\langle j|\psi _{l}\rangle \langle \psi
_{l}|j\rangle \geq 0$ and $\rho _{jj}\geq 0,$ then Eq. (\ref{eq2.1}) obviously holds.
Suppose the pure state decomposition $\rho =\sum_{l}p_{l}|\psi _{l}\rangle
\langle \psi _{l}|$ achieves $\overline{C_{l_{1}}}(\rho )=C_{l_{1}}(\rho ),$
i.e., $C_{l_{1}}(\rho )=\sum_{l}p_{l}C_{l_{1}}(|\psi _{l}\rangle \langle
\psi _{l}|).$ $\rho =\sum_{l}p_{l}|\psi _{l}\rangle \langle \psi _{l}|$
implies
\begin{eqnarray}
\langle j|\rho |k\rangle &=&\sum_{l}p_{l}\langle j|\psi _{l}\rangle \langle
\psi _{l}|k\rangle ,  \label{eq2.2} \\
|\langle j|\rho |k\rangle | &=&|\sum_{l}p_{l}\langle j|\psi _{l}\rangle
\langle \psi _{l}|k\rangle |\leq \sum_{l}p_{l}|\langle j|\psi _{l}\rangle
\langle \psi _{l}|k\rangle |,  \label{eq2.3} \\
\sum_{j\neq k}|\langle j|\rho |k\rangle | &=&\sum_{j\neq
k}|\sum_{l}p_{l}\langle j|\psi _{l}\rangle \langle \psi _{l}|k\rangle |\leq
\sum_{j\neq k}\sum_{l}p_{l}|\langle j|\psi _{l}\rangle \langle \psi
_{l}|k\rangle |, \label{eq2.4}
\end{eqnarray}
$C_{l_{1}}(\rho )=\sum_{l}p_{l}C_{l_{1}}(|\psi _{l}\rangle \langle \psi
_{l}|)$ implies
\begin{eqnarray}
\sum_{j\neq k}|\langle j|\rho |k\rangle |=\sum_{j\neq
k}\sum_{l}p_{l}|\langle j|\psi _{l}\rangle \langle \psi _{l}|k\rangle |. \label{eq2.5}
\end{eqnarray}

We know that for complex numbers $\{z_{j}\}_{j},$ $|\sum_{j}z_{j}|=%
\sum_{j}|z_{j}|$ if and only if there exists a constant real number $\theta $
such that $z_{j}=|z_{j}|e^{i\theta }$ for all $j$. With this fact, we then
obtain Eq. (\ref{eq2.1}). $\hfill\blacksquare$

We provide three lemmas below which will be useful in following
sections.

\emph{Lemma 1.} For any coherence measure $C,$ any state $\rho $ and any incoherent
unitary transformation $U$ in $d$-dimensional system,
\begin{eqnarray}
U=\left(
\begin{array}{cccc}
e^{i\theta _{1}} & 0 & ... & 0 \\
0 & e^{i\theta _{2}} & ... & 0 \\
... & ... & ... & ... \\
0 & 0 & ... & e^{i\theta _{d}}%
\end{array}%
\right) ,  \label{eq2.6}
\end{eqnarray}
it holds that $U$ preserves $C(\rho )$, i.e., $C(\rho )=C(U\rho U^{\dagger
})$. Where $\{\theta _{j}\}_{j=1}^{d}$ are all real numbers. Further, $\overline{C}(\rho )=C(\rho )$ if and only if $\overline{C}(U\rho U^{\dagger})=C(U\rho U^{\dagger})$.

\emph{Proof.} We can check that $U$ and $U^{\dagger }$ are all incoherent channels,
i.e., $U\sigma U^{\dagger }$ and $U^{\dagger }\sigma U$ are diagonal for any
diagonal state $\sigma .$ From the monotonicity (C2) of coherence measure,
we have $C(U\rho U^{\dagger })\leq C(\rho )=C(U^{\dagger }U\rho U^{\dagger
}U)\leq C(U\rho U^{\dagger }).$ Hence $C(\rho )=C(U\rho U^{\dagger})$.
For the pure state decomposition $\rho
=\sum_{l}p_{l}|\psi _{l}\rangle \langle \psi _{l}|,$ we have $U\rho U^{\dagger }
=\sum_{l}p_{l}U|\psi _{l}\rangle \langle \psi _{l}|U^{\dagger };$ if further $C(\rho
)=\sum_{l}p_{l}C(|\psi _{l}\rangle \langle \psi _{l}|)$ then $C(U\rho
U^{\dagger })=\sum_{l}p_{l}C(U|\psi _{l}\rangle \langle \psi _{l}|U^{\dagger
}).$ That is to say, $\overline{C}(\rho )=C(\rho )$ if and only if $\overline{C}(U\rho U^{\dagger
})=C(U\rho U^{\dagger})$. $\hfill\blacksquare$

\emph{Lemma 2.} (See Corollary 7.1.5 and Theorem 7.2.5 in Ref. \cite{Horn-2013-book}%
) Let $A$ be a Hermitian matrix, then $A$ is positive semidefinite if and
only if every principal minor of $A$ (including det$A$) is nonnegative.

A matrix $A$ is called a positive (nonnegative) matrix if all its elements are positive (nonnegative).
For positive matrix, we have Lemma 3 below.

\emph{Lemma 3.} (See Theorem 8.2.8 in Ref. \cite{Horn-2013-book}). For the $d$%
-dimensional positive matrix $A,$ the largest eigenvalue of $A,$ $\lambda
_{\max }(A),$ is positive and algebraically simple, and corresponds to a
normalized eigenvector $|\chi \rangle =\sum_{j=1}^{d}\chi _{j}|j\rangle $
with $\{\chi _{j}>0\}_{j=1}^{d}.$

\section{$\overline{C_{l_{1}}}=C_{l_{1}}$ for $d=2$}

In Ref. \cite{Qi-2017-JPA}, it is shown that $\overline{C_{l_{1}}}=C_{l_{1}}$
for $d=2.$ We derive this result in a different way. For any qubit state $%
\rho ,$ we write it in the Bloch representation as
\begin{eqnarray}
\rho =\frac{1}{2}(I_{2}+\overrightarrow{r}\cdot \overrightarrow{\sigma })=%
\frac{1}{2}\left(
\begin{array}{cc}
1+z & x-iy \\
x+iy & 1-z%
\end{array}%
\right) ,   \label{eq3.1}
\end{eqnarray}
with $\overrightarrow{r}=(x,y,z)$ real vector, $r=|\overrightarrow{r}|\leq
1$, and $\overrightarrow{\sigma }=(\sigma_{x},\sigma_{y},\sigma_{z})$ the Pauli matrices.

A pure state decomposition $\rho =\sum_{l}p_{l}|\psi _{l}\rangle \langle
\psi _{l}|$ can be expressed in the Bloch representation as
\begin{eqnarray}
\rho =\frac{1}{2}(I_{2}+\sum_{l}p_{l}\overrightarrow{n_{l}}\cdot
\overrightarrow{\sigma }),  \label{eq3.2}
\end{eqnarray}
where $|\psi _{l}\rangle \langle \psi _{l}|=\frac{1}{2}(I_{2}+%
\overrightarrow{n_{l}}\cdot \overrightarrow{\sigma })$, $\overrightarrow{%
n_{l}}=(n_{lx},n_{ly},n_{lz})$ real vector, $|\overrightarrow{n_{l}}|=1$, and
\begin{eqnarray}
\sum_{l}p_{l}\overrightarrow{n_{l}}=\overrightarrow{r}. \label{eq3.3}
\end{eqnarray}

Suppose $\rho =\sum_{l}p_{l}|\psi _{l}\rangle \langle \psi _{l}|$ achieves $%
\overline{C_{l_{1}}}(\rho )=C_{l_{1}}(\rho )$, applying Theorem 1, we get
that there exist $\{\mu _{l}\geq 0\}_{l}$ such that
\begin{eqnarray}
\overrightarrow{n_{l}}=(\mu _{l}x,\mu _{l}y,n_{lz}),\forall l. \label{eq3.4}
\end{eqnarray}
We stress that when $r=1$, $\rho (x,y,z)$ is pure, we have $\overline{%
C_{l_{1}}}(\rho )=C_{l_{1}}(\rho ).$ When $\rho (x,y,z)=\rho (0,0,z)=\frac{1%
}{2}\left(
\begin{array}{cc}
1+z & 0 \\
0 & 1-z%
\end{array}%
\right) ,$ we see that $\overline{C_{l_{1}}}(\rho )=0$ and $\frac{I_{2}}{2}=%
\frac{1+z}{2}|1\rangle \langle 1|+\frac{1-z}{2}|2\rangle \langle 2|$
achieves $\overline{C_{l_{1}}}(\rho )=C_{l_{1}}(\rho ).$ Then we only
consider mixed states $|z|<r<1.$

We depict Eqs. (\ref{eq3.3},\ref{eq3.4}) in Figure 2(a). Suppose the state $\rho (x,y,z)$ with $%
r\in (0,1)$ and $\sqrt{x^{2}+y^{2}}>0,$ corresponds to the point $E(x,y,z)$
in the Bloch representation. The points $O(0,0,0)$, $D(0,0,1)$, and $E(x,y,z)
$ determine a plane ODE, this plane intersects the x-O-y plane at the points
$A(-\frac{x}{\sqrt{x^{2}+y^{2}}},-\frac{y}{\sqrt{x^{2}+y^{2}}},0)$, $B(\frac{%
x}{\sqrt{x^{2}+y^{2}}},\frac{y}{\sqrt{x^{2}+y^{2}}},0)$. If $\rho
=\sum_{l=1}^{m}p_{l}|\psi _{l}\rangle \langle \psi _{l}|$ achieves $%
\overline{C_{l_{1}}}(\rho )=C_{l_{1}}(\rho ),$ this fact corresponds to that
there is a polygonal Line $OO_{1}O_{2}...O_{m}$ in ODE plane such that $%
\overrightarrow{O_{l-1}O_{l}}=p_{l}\overrightarrow{n_{l}}$ with $O_{0}=O$, $%
O_{m}=E$ and $\overrightarrow{n_{l}}=(\mu _{l}x,\mu _{l}y,n_{lz})$, $\{\mu
_{l}\geq 0\}_{l}.$ From Figure 2(a), we see that there is an infinite number
of pure state decompositions $\rho =\sum_{l=1}^{m}p_{l}|\psi _{l}\rangle
\langle \psi _{l}|$ achieving $\overline{C_{l_{1}}}(\rho )=C_{l_{1}}(\rho ).$
\begin{figure}[!ht]
\centering
  \begin{minipage}[t]{0.34\textwidth}
    \centering
    \includegraphics[width=1\textwidth,bb=85 400 500 750]{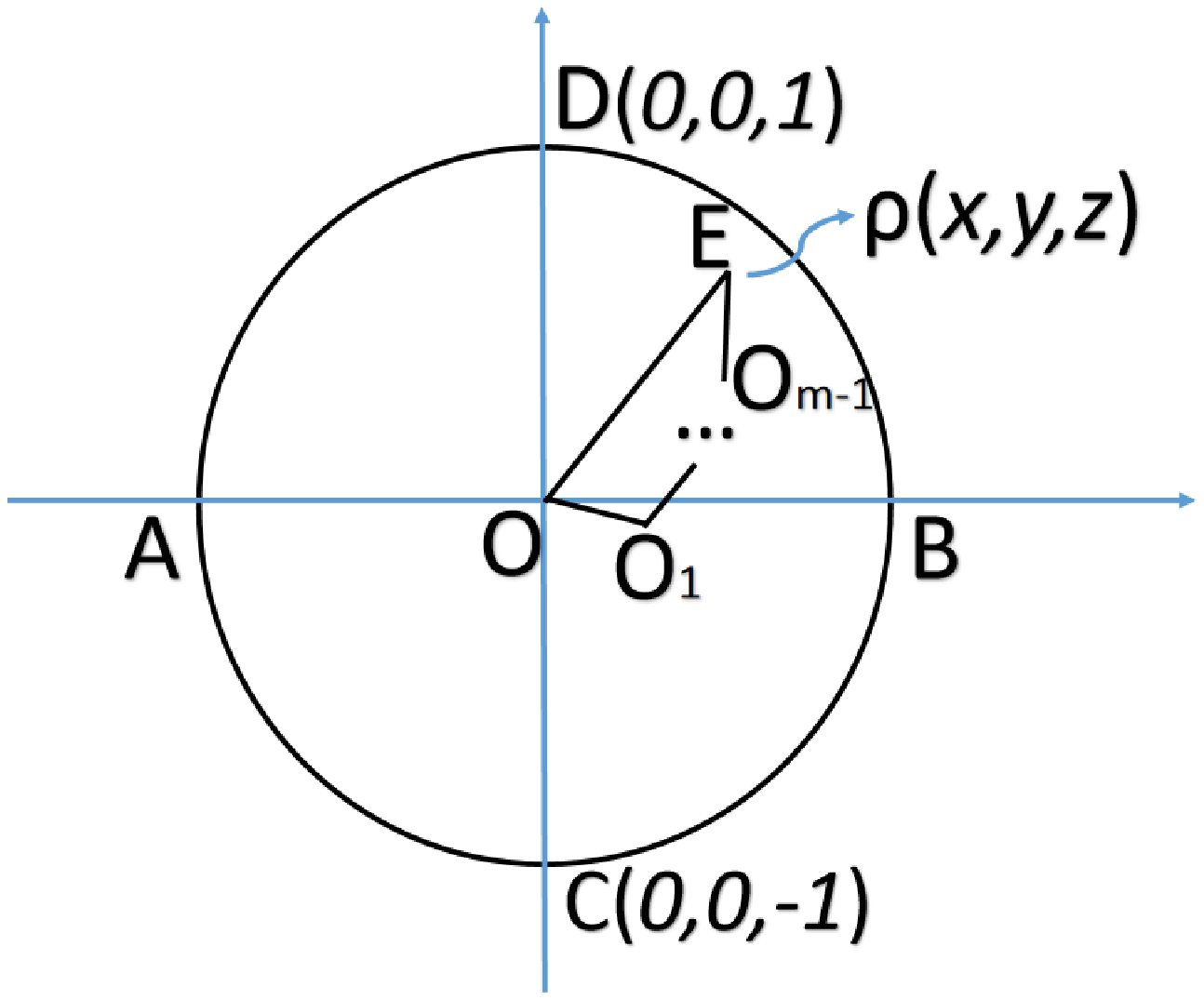}
    Fig.2(a):Depiction of Eqs. (\ref{eq3.3},\ref{eq3.4})
  \end{minipage}
  \begin{minipage}[t]{0.30\textwidth}
    \centering
    \includegraphics[width=1\textwidth,bb=100 412 510 800]{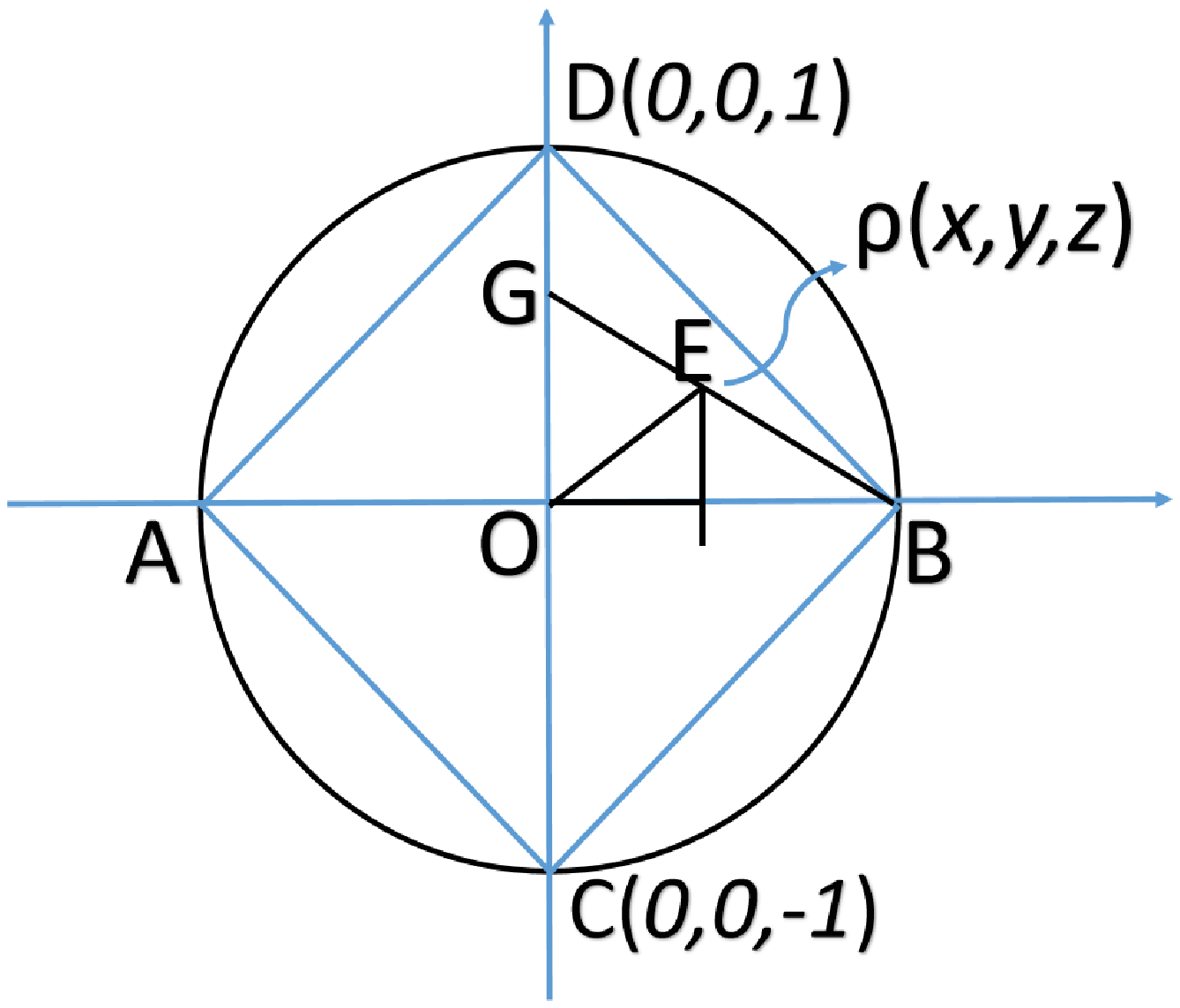}
    Fig.2(b):$\sqrt{x^{2}+y^{2}}+|z|\leq 1$
  \end{minipage}
  \begin{minipage}[t]{0.34\textwidth}
    \centering
    \includegraphics[width=1\textwidth,bb=100 440 500 770]{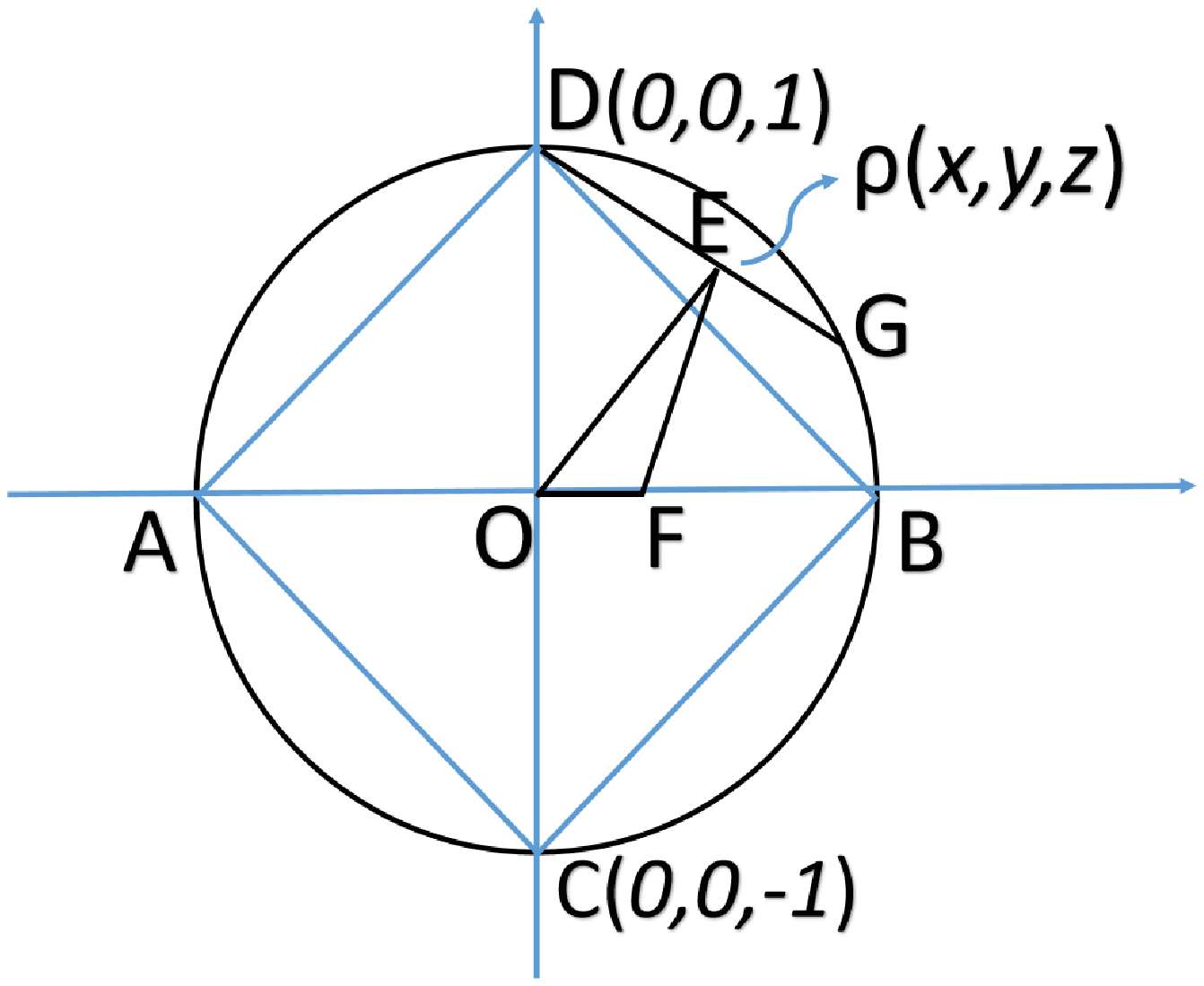}
    Fig.2(c):$\sqrt{x^{2}+y^{2}}+|z|\leq 1$
  \end{minipage}
\end{figure}

We provide an explicit pure state decomposition $\rho =\sum_{l}p_{l}|\psi
_{l}\rangle \langle \psi _{l}|$ achieving $\overline{C_{l_{1}}}(\rho
)=C_{l_{1}}(\rho )$ as follows. When $\sqrt{x^{2}+y^{2}}+|z|\leq 1,$ let
\begin{eqnarray}
p_{1}\overrightarrow{n_{1}} &=&(x,y,0),  \label{eq3.5} \\
p_{2}\overrightarrow{n_{2}} &=&(0,0,-sz),  \label{eq3.6} \\
p_{3}\overrightarrow{n_{3}} &=&(0,0,(1+s)z),  \label{eq3.7} \\
0 \leq s&=&\frac{1}{2}(\frac{1-\sqrt{x^{2}+y^{2}}}{|z|}-1). \label{eq3.8}
\end{eqnarray}%
We can check that this pure state decomposition achieves $\overline{C_{l_{1}}%
}(\rho )=C_{l_{1}}(\rho ),$ we depict this pure state decomposition in
Figure 2(b). \ When $\sqrt{x^{2}+y^{2}}+|z|\geq 1,$ let
\begin{eqnarray}
p_{1}\overrightarrow{n_{1}} &=&s(x,y,0),  \label{eq3.9} \\
p_{2}\overrightarrow{n_{2}} &=&((1-s)x,(1-s)y,z),  \label{eq3.10} \\
0 \leq s &=&\frac{1-r^{2}}{2\sqrt{x^{2}+y^{2}}(1-\sqrt{x^{2}+y^{2}})}. \label{eq3.11}
\end{eqnarray}%
We can check that this pure state decomposition achieves $\overline{C_{l_{1}}%
}(\rho )=C_{l_{1}}(\rho ),$ we depict this pure state decomposition in
Figure 2(c).

\section{$\overline{C_{l_{1}}}\neq C_{l_{1}}$ for $d=3$}

For $d=3$, we consider the states in different situations.

(1). $\rho _{11}\rho _{22}\rho _{33}=0.$ For this case, at least one of $%
\{\rho _{jj}\}_{j=1}^{3}$ is zero. For example $\rho _{11}=0,$ from Lemma 2
we then have  $\{\rho _{1j}=0\}_{j=1}^{3}$ and $\{\rho _{j1}=0\}_{j=1}^{3}$,
$\rho $ degenerates to a $2$-dimensional state, from the fact $\overline{C_{l_{1}}}=C_{l_{1}}$ for $d=2$ obtained in section III we then
have $C_{l_{1}}(\rho )=\overline{C_{l_{1}}}(\rho ).$

(2). $\rho _{11}\rho _{22}\rho _{33}>0,\Pi _{j\neq k}\rho _{jk}=0.$ For this
case, at least one of $\{\rho _{jk}\}_{j\neq k}$ is zero. For example $\rho
_{13}=0,$ then
\begin{eqnarray}
\rho =\left(
\begin{array}{ccc}
\rho _{11} & |\rho _{12}|e^{i\theta _{12}} & 0 \\
|\rho _{12}|e^{-i\theta _{12}} & \rho _{22} & |\rho _{23}|e^{i\theta _{23}}
\\
0 & |\rho _{23}|e^{-i\theta _{23}} & \rho _{33}%
\end{array}%
\right).   \label{eq4.1}
\end{eqnarray}
If further $\rho _{12}\rho _{23}=0,$ then $\rho $ can be expressed as a
direct sum of two states with dimensions at most $2$. According to (C5) and $%
\overline{C_{l_{1}}}=C_{l_{1}}$ for $d=2$, we get that $\overline{C_{l_{1}}}%
(\rho )=C_{l_{1}}(\rho ).$
So we assume $|\rho _{11}\rho _{22}\rho _{33}\rho _{12}\rho _{23}|>0.$

We decompose $\rho $ as
\begin{eqnarray}
\rho &=&p\rho _{1}+(1-p)\rho _{2},   \label{eq4.2} \\
p\rho _{1} &=&\left(
\begin{array}{ccc}
\rho _{11} & |\rho _{12}|e^{i\theta _{12}} & 0   \\
|\rho _{12}|e^{-i\theta _{12}} & \frac{|\rho _{12}|^{2}}{\rho _{11}} & 0 \\
0 & 0 & 0%
\end{array}%
\right) ,   \label{eq4.3} \\
(1-p)\rho _{2} &=&\left(
\begin{array}{ccc}
0 & 0 & 0 \\
0 & \rho _{22}-\frac{|\rho _{12}|^{2}}{\rho _{11}} & |\rho _{23}|e^{i\theta
_{23}} \\
0 & |\rho _{23}|e^{-i\theta _{23}} & \rho _{33}%
\end{array}%
\right) ,   \label{eq4.4}
\end{eqnarray}
with $p\in (0,1)$ such that tr$\rho _{1}=$tr$\rho _{2}=$tr$\rho =1.$ Notice
that for $(1-p)\rho _{2}$ and $\rho ,$ because
\begin{eqnarray}
\text{det}\rho =\rho _{11}\rho _{22}\rho _{33}-|\rho _{12}|^{2}\rho
_{33}-|\rho _{23}|^{2}\rho _{11}\geq 0,   \label{eq4.5}
\end{eqnarray}
then
\begin{eqnarray}
\text{det}\left(
\begin{array}{cc}
\rho _{22}-\frac{|\rho _{12}|^{2}}{\rho _{11}} & |\rho _{23}|e^{i\theta
_{23}} \\
|\rho _{23}|e^{-i\theta _{23}} & \rho _{33}%
\end{array}%
\right) =(\rho _{22}-\frac{|\rho _{12}|^{2}}{\rho _{11}})\rho _{33}-|\rho
_{23}|^{2}\geq 0,   \label{eq4.6}
\end{eqnarray}
and $(1-p)\rho _{2}$ is positive semidefinite. Using Lemma 1, we get that $%
p\rho _{1}$ and $(1-p)\rho _{2}$ are all positive semidefinite, and then $%
\rho _{1}$ and $\rho _{2}$ are all quantum states. Since $\overline{C_{l_{1}}%
}=C_{l_{1}}$ for $d=2$, we then get that $\overline{C_{l_{1}}}(\rho
)=C_{l_{1}}(\rho ).$

(3). $\Pi _{j,k=1}^{3}|\rho _{jk}|>0,\theta _{23}=\theta _{13}-\theta _{12}.$
For this case,
\begin{eqnarray}
\rho =\left(
\begin{array}{ccc}
\rho _{11} & |\rho _{12}|e^{i\theta _{12}} & |\rho _{13}|e^{i\theta _{13}}
\\
|\rho _{12}|e^{-i\theta _{12}} & \rho _{22} & |\rho _{23}|e^{i(\theta
_{13}-\theta _{12})} \\
|\rho _{13}|e^{-i\theta _{13}} & |\rho _{23}|e^{-i(\theta _{13}-\theta
_{12})} & \rho _{33}%
\end{array}%
\right) .    \label{eq4.7}
\end{eqnarray}

Let the unitary matrix $U$ be
\begin{eqnarray}
U=\left(
\begin{array}{ccc}
1 & 0 & 0 \\
0 & e^{i\theta _{12}} & 0 \\
0 & 0 & e^{i\theta _{13}}%
\end{array}%
\right) .   \label{eq4.8}
\end{eqnarray}
Then
\begin{eqnarray}
U\rho U^{\dagger }=\left(
\begin{array}{ccc}
\rho _{11} & |\rho _{12}| & |\rho _{13}| \\
|\rho _{12}| & \rho _{22} & |\rho _{23}| \\
|\rho _{13}| & |\rho _{23}| & \rho _{33}%
\end{array}%
\right) .   \label{eq4.9}
\end{eqnarray}
We see that all elements $\{|\rho _{jk}|\}_{jk=1}^{3}$ of $U\rho U^{\dagger }
$ are positive, then $U\rho U^{\dagger }$ is a positive matrix and also a
positive semidefinite matrix.

Since $U\rho U^{\dagger }$ and $\rho $ have the same eigenvalues, applying
Lemma 3, we get that $U\rho U^{\dagger }$ has the largest eigenvalue $\lambda
_{\max }(U\rho U^{\dagger })=\lambda _{\max }(\rho )$ which corresponds to
the positive normalized eigenvector $|\chi \rangle =\sum_{j=1}^{3}\chi
_{j}|j\rangle $ with $\{\chi _{j}>0\}_{j=1}^{3}.$ Consider the state
\begin{eqnarray}
\rho (t)=U\rho U^{\dagger }-t|\chi \rangle \langle \chi |.   \label{eq4.10}
\end{eqnarray}
We see that $|\chi \rangle \langle \chi |$ is positive and positive
semidefinite, $\rho (0)=U\rho U^{\dagger }$ is positive and positive
semidefinite, $\rho (\lambda _{\max }(\rho ))$ is positive semidefinite.

If $\rho (\lambda _{\max }(\rho ))$ is no longer positive, then there must
exist $0<t_{0}<\lambda _{\max }(\rho )$ such that $\rho (t_{0})=U\rho
U^{\dagger }-t_{0}|\chi \rangle \langle \chi |$ is nonnegative and at least one element $%
\langle j|\rho (t_{0})|k\rangle =0.$ Hence $\rho (t_{0})$ turns to situation
(1) or (2), and there exists a pure state decomposition $U\rho U^{\dagger
}-t_{0}|\chi \rangle \langle \chi |=\sum_{l}q_{l}|\psi _{l}\rangle \langle
\psi _{l}|$ such that $\{q_{l}>0\}_{l}$, $\sum_{l}q_{l}=1-t_{0}$, $\{|\psi
_{l}\rangle >0\}_{l}$ are all normalized pure states and $\{\langle j|\psi
_{l}\rangle \langle \psi _{l}|k\rangle \geq 0\}_{l,j,k}$. Employing Theorem
1 and Lemma 1, we see that $U\rho U^{\dagger }=t_{0}|\chi \rangle \langle
\chi |+\sum_{l}q_{l}|\psi _{l}\rangle \langle \psi _{l}|$ is a pure state
decomposition which achieves $C_{l_{1}}(U\rho U^{\dagger } )=t_{0}C_{l_{1}}(|\chi \rangle
\langle \chi |)+\sum_{l}q_{l}C_{l_{1}}(|\psi _{l}\rangle \langle \psi _{l}|).
$ That is to say, $\overline{C_{l_{1}}}(U\rho U^{\dagger })=C_{l_{1}}(U\rho U^{\dagger } )$ and $\overline{C_{l_{1}}}(\rho )=C_{l_{1}}(\rho ).$

If $\rho (\lambda _{\max }(\rho ))$ is still positive, then let
\begin{eqnarray}
\rho _{1}(t)=U\rho U^{\dagger }-\lambda _{\max }(\rho )|\chi \rangle \langle
\chi |-t^{\prime }|\chi ^{\prime }\rangle \langle \chi ^{\prime }|,   \label{eq4.11}
\end{eqnarray}
where $|\chi ^{\prime }\rangle $ is the normalized eigenvector associated
with the largest eigenvalue of $\rho (\lambda _{\max }(\rho )),$ denoted by $%
\lambda _{\max }(\rho (\lambda _{\max }(\rho ))).$ Discuss $\rho _{1}(t)$
similarly to $\rho (t)$, and repeat this process with a finite number of
repetitions, we will finally arrive at $\overline{C_{l_{1}}}(\rho
)=C_{l_{1}}(\rho ).$

(4). $\Pi _{jk=1}^{3}|\rho _{jk}|>0,\theta _{23}\neq \theta
_{13}-\theta _{12}.$ For this case,
\begin{eqnarray}
\rho =\left(
\begin{array}{ccc}
\rho _{11} & |\rho _{12}|e^{i\theta _{12}} & |\rho _{13}|e^{i\theta _{13}}
\\
|\rho _{12}|e^{-i\theta _{12}} & \rho _{22} & |\rho _{23}|e^{i\theta _{23}}
\\
|\rho _{13}|e^{-i\theta _{13}} & |\rho _{23}|e^{-i\theta _{23}} & \rho _{33}%
\end{array}%
\right) .  \label{eq4.12}
\end{eqnarray}
Let the unitary matrix $U$ be
\begin{eqnarray}
U=\left(
\begin{array}{ccc}
1 & 0 & 0 \\
0 & e^{i\theta _{12}} & 0 \\
0 & 0 & e^{i\theta _{13}}%
\end{array}%
\right) . \label{eq4.13}
\end{eqnarray}
Then
\begin{eqnarray}
U\rho U^{\dagger }=\left(
\begin{array}{ccc}
\rho _{11} & |\rho _{12}| & |\rho _{13}| \\
|\rho _{12}| & \rho _{22} & |\rho _{23}|e^{i\theta } \\
|\rho _{13}| & |\rho _{23}|e^{-i\theta } & \rho _{33}%
\end{array}%
\right) ,  \label{eq4.14}
\end{eqnarray}
with $\theta =\theta _{12}+\theta _{23}-\theta _{13}$ and $e^{i\theta }\neq
1.$

Suppose the pure state decomposition $U\rho U^{\dagger }=\sum_{l}p_{l}|\psi
_{l}\rangle \langle \psi _{l}|$ achieves $\overline{C_{l_{1}}}(U\rho
U^{\dagger })=C_{l_{1}}(U\rho U^{\dagger }).$ Let $|\psi _{l}\rangle
=\sum_{j=1}^{3}\psi _{lj}|j\rangle $ with $\psi _{lj}=|\psi _{lj}|e^{i\theta
_{l,j}}$ its polar form, then
\begin{eqnarray}
|\psi _{l}\rangle \langle \psi _{l}|=\sum_{j,k=1}^{3}|\psi _{lj}\psi
_{lk}|e^{i(\theta _{l,j}-\theta _{l,k})}.  \label{eq4.15}
\end{eqnarray}

If $|\psi _{l1}\psi _{l2}\psi _{l3}|>0,$ applying Theorem 1, Eq. (\ref{eq4.15}) and the
first row of $U\rho U^{\dagger }$ imply $\theta _{l,1}=\theta _{l,2}=\theta
_{l,3}$ while Eq. (\ref{eq4.15}) and the second row of $U\rho U^{\dagger }$ imply $%
\theta _{l,1}=\theta _{l,2}\neq \theta _{l,3}.$ This contradiction implies
that $|\psi _{l1}\psi _{l2}\psi _{l3}|=0,$ and then any $|\psi _{l}\rangle
\langle \psi _{l}|$ of $\{|\psi _{l}\rangle \langle \psi _{l}|\}_{l}$ has
one of the forms
\begin{eqnarray}
|\psi _{l}\rangle \langle \psi _{l}| &=&\left(
\begin{array}{ccc}
|\psi _{l1}|^{2} & |\psi _{l1}\psi _{l2}| & 0 \\
|\psi _{l1}\psi _{l2}| & |\psi _{l2}|^{2} & 0 \\
0 & 0 & 0%
\end{array}%
\right) ,\psi _{l3}=0,  \label{eq4.16}  \\
|\psi _{l}\rangle \langle \psi _{l}| &=&\left(
\begin{array}{ccc}
0 & 0 & 0 \\
0 & |\psi _{l2}|^{2} & |\psi _{l2}\psi _{l3}|e^{i\theta } \\
0 & |\psi _{l2}\psi _{l3}|e^{-i\theta } & |\psi _{l3}|^{2}%
\end{array}%
\right) ,\psi _{l1}=0,  \label{eq4.17}  \\
|\psi _{l}\rangle \langle \psi _{l}| &=&\left(
\begin{array}{ccc}
|\psi _{l1}|^{2} & 0 & |\psi _{l1}\psi _{l3}| \\
0 & 0 & 0 \\
|\psi _{l1}\psi _{l3}| & 0 & |\psi _{l3}|^{2}%
\end{array}%
\right) ,\psi _{l2}=0.  \label{eq4.18}
\end{eqnarray}

Consequently, $U\rho U^{\dagger }=\sum_{l}p_{l}|\psi _{l}\rangle \langle
\psi _{l}|$ achieving $\overline{C_{l_{1}}}(\rho )=C_{l_{1}}(\rho )$ results
in the decomposition
\begin{eqnarray}
U\rho U^{\dagger }=\left(
\begin{array}{ccc}
x_{1} & |\rho _{12}| & 0 \\
|\rho _{12}| & x_{2} & 0 \\
0 & 0 & 0%
\end{array}%
\right) +\left(
\begin{array}{ccc}
0 & 0 & 0 \\
0 & \rho _{22}-x_{2} & |\rho _{23}|e^{i\theta } \\
0 & |\rho _{23}|e^{-i\theta } & x_{3}%
\end{array}%
\right) +\left(
\begin{array}{ccc}
\rho _{11}-x_{1} & 0 & |\rho _{13}| \\
0 & 0 & 0 \\
|\rho _{13}| & 0 & \rho _{33}-x_{3}%
\end{array}%
\right) ,  \label{eq4.19}
\end{eqnarray}
with $0<x_{1}<\rho _{11},0<x_{2}<\rho _{22},0<x_{3}<\rho _{33}$, and each
matrix of the right side of Eq. (\ref{eq4.19}) is positive definite. Since the first
matrix of the right side of Eq. (\ref{eq4.19}) is positive definite, then $%
x_{1}x_{2}\geq |\rho _{12}|^{2},$ and we decompose
\begin{eqnarray}
\left(
\begin{array}{ccc}
x_{1} & |\rho _{12}| & 0 \\
|\rho _{12}| & x_{2} & 0 \\
0 & 0 & 0%
\end{array}%
\right) =\left(
\begin{array}{ccc}
x_{1} & |\rho _{12}| & 0 \\
|\rho _{12}| & \frac{|\rho _{12}|^{2}}{x_{1}} & 0 \\
0 & 0 & 0%
\end{array}%
\right) +\left(
\begin{array}{ccc}
0 & 0 & 0 \\
0 & x_{2}-\frac{|\rho _{12}|^{2}}{x_{1}} & 0 \\
0 & 0 & 0%
\end{array}%
\right) .  \label{eq4.20}
\end{eqnarray}
As a result, we get
\begin{eqnarray}
U\rho U^{\dagger }=\left(
\begin{array}{ccc}
x_{1} & |\rho _{12}| & 0 \\
|\rho _{12}| & \frac{|\rho _{12}|^{2}}{x_{1}} & 0 \\
0 & 0 & 0%
\end{array}%
\right) +\overline{\rho },  \label{eq4.21}  \\
\overline{\rho }=\left(
\begin{array}{ccc}
\rho _{11}-x_{1} & 0 & |\rho _{13}| \\
0 & \rho _{22}-\frac{|\rho _{12}|^{2}}{x_{1}} & |\rho _{23}|e^{i\theta } \\
|\rho _{13}| & |\rho _{23}|e^{-i\theta } & \rho _{33}%
\end{array}%
\right) .  \label{eq4.22}
\end{eqnarray}

Combining with situation (2), we see that $\overline{C_{l_{1}}}(\rho
)=C_{l_{1}}(\rho )$ if and only if there exists $0<x_{1}<\rho _{11}$ such
that $\overline{\rho }$ is positive definite. Now we exemplify that for some
states $\overline{C_{l_{1}}}(\rho )=C_{l_{1}}(\rho ),$ but for some states $%
\overline{C_{l_{1}}}(\rho )>C_{l_{1}}(\rho ).$

\emph{Example 1.} Let
\begin{eqnarray}
\rho (|\rho _{13}|) &=&\left(
\begin{array}{ccc}
0.1 & 0.01 & |\rho _{13}| \\
0.01 & 0.1 & 0.2i \\
|\rho _{13}| & -0.2i & 0.8%
\end{array}%
\right) =\left(
\begin{array}{ccc}
x_{1} & 0.01 & 0 \\
0.01 & \frac{0.0001}{x_{1}} & 0 \\
0 & 0 & 0%
\end{array}%
\right) +\overline{\rho (|\rho _{13}|)},  \label{eq4.23} \\
\overline{\rho (|\rho _{13}|)} &=&\left(
\begin{array}{ccc}
0.1-x_{1} & 0 & |\rho _{13}| \\
0 & 0.1-\frac{0.0001}{x_{1}} & 0.2i \\
|\rho _{13}| & 0.2i & 0.8%
\end{array}%
\right) .  \label{eq4.24}
\end{eqnarray}

When $|\rho _{13}|=0.17,$ with direct calculation, we get that the
eigenvalues of $\rho (|\rho _{13}|=0.17)$ approximate $%
\{0.887506,0.101004,0.0114902\}.$ Then $\rho (|\rho _{13}|=0.17)$ is positive semidefinite and is a density matrix. Let $x_{1}=0.0113,$ the eigenvalues of $%
\overline{\rho (|\rho _{13}|=0.17)}$ approximate $%
\{0.886515,0.089743,0.00359217\}.$ Then when $x_{1}=0.0113,$ $\overline{\rho
(|\rho _{13}|=0.17)}$ is positive definite, and $\overline{C_{l_{1}}}(\rho
(|\rho _{13}|=0.17))=C_{l_{1}}(\rho (|\rho _{13}|=0.17)).$

When $|\rho _{13}|=0.19,$ with direct calculation, we get that the
eigenvalues of $\rho (|\rho _{13}|=0.19)$ approximate $%
\{0.895659,0.100911,0.00342989\}.$ Then $\rho (|\rho _{13}|=0.19)$ is positive semidefinite and is a density matrix. We approximately calculate det$\overline{\rho (|\rho _{13}|=0.19)}$ as
\begin{eqnarray}
&&\text{det}\overline{\rho (|\rho _{13}|=0.19)} \\
&\approx &0.00047-(\frac{4.39\times 10^{-6}}{x_{1}}+0.04x_{1}) \\
&\leq &0.00047-2\sqrt{4.39\times 10^{-6}\times 0.04} \\
&\approx &-3.68\times 10^{-4},
\end{eqnarray}
in the inequality we have used $a+b\geq 2\sqrt{ab}$ for $a>0$ and $b>0.$
This shows $\overline{\rho (|\rho _{13}|=0.19)}$ is not positive
semidefinite for any $0<x_{1}<0.1,$ then $\overline{C_{l_{1}}}(\rho (|\rho
_{13}|=0.19))>C_{l_{1}}(\rho (|\rho _{13}|=0.19)).$ This example shows $%
\overline{C_{l_{1}}}(\rho )\neq C_{l_{1}}(\rho ).$

\section{Summary}
For a given coherence measure $C$, extending the values of $C$ on pure states to mixed states by the convex roof construction, we will get a valid coherence measure $\overline{C}$. 
In this work, we showed that $\overline{C_{l_{1}}}(\rho )\neq C_{l_{1}}(\rho
)$ for the widely used coherence measure, $l_{1}$ norm of coherence $C_{l_{1}}$.  To this aim, we first established a criterion for $\overline{C_{l_{1}}}(\rho
)=C_{l_{1}}(\rho )$ which lays a foundation for the subsequent discussions.
We proved $\overline{C_{l_{1}}}(\rho )=C_{l_{1}}(\rho )$ for $d=2$ and
provided a geometric interpretation in Bloch representation. We
investigated $\overline{C_{l_{1}}}(\rho )$ for $d=3$ in different
situations, in three situations we proved $\overline{C_{l_{1}}}(\rho
)=C_{l_{1}}(\rho )$ and in last situation we showed that both $\overline{%
C_{l_{1}}}(\rho )=C_{l_{1}}(\rho )$ and $\overline{C_{l_{1}}}(\rho
)>C_{l_{1}}(\rho )$ are all possible. An explicit example is given to show $%
\overline{C_{l_{1}}}(\rho )\neq C_{l_{1}}(\rho ).$

\section*{ACKNOWLEDGMENTS}
This work was supported by the Chinese Universities Scientific Fund under Grant No. 2452021067.


\bibliographystyle{apsrev4-1}
%

\end{document}